# Electronic structure for new layered high-temperature superconductors $CaAFe_4As_4$ (A=K, Rb, Cs): FLAPW-GGA calculations

*D.V. Suetin, I.R. Shein\**

*Institute of Solid State Chemistry, Ural Branch of the Russian Academy of Sciences, Ekaterinburg, 620990, Russia*

Recently, new FeAs based high-temperature superconductors $CaAFe_4As_4$ (A=K, Rb, Cs) with a layered tetragonal crystal structure were synthesized ($T_C \sim 30$ K). In this Letter, we report for the first time the band structures, Fermi surface topology, total and partial densities of electronic states and interatomic interactions for $CaAFe_4As_4$ as estimated by means of the first-principles FLAPW-GGA calculations. The interatomic bonding picture can be represented as a highly anisotropic mixture of metallic, covalent, and ionic contributions, which are realized inside $Fe_4As_4$ layered blocks and between these blocks and Ca, A atomic sheets. The Fermi surfaces of these systems have a multisheet character and are compiled of a large number of cylinders at the edges and in the central part of the Brillouin zone. It is established that the high-temperature superconductivity in $CaAFe_4As_4$ compounds as in other related systems correlates well with such structure parameters as bond angles and anion height.

*Keywords:* High-temperature superconductors $CaAFe_4As_4$, Electronic structure; Fermi surface; Interatomic interactions; First-principles calculations

[*] Corresponding author.
*E-mail address:* shein@ihim.uran.ru (I.R. Shein).

**I. Introduction.** Since the discovery of unconventional superconductivity which is not related to electron-phonon pairing in F doped La-Fe oxyarsenide LaFeAsO$_{1-x}$F$_x$ with $T_C$=26 K [1], much attention was paid to the search for new superconductors based on systems with FePn blocks or related FeCh blocks, where Pn is a pnictogen atom and Ch is a chalcogen atom. Such properties can be explained by the appearance of magnetic fluctuations or spin and charge density waves (SDW or CDW). Except for superconductivity, these systems possess a number of remarkable physical properties such as magnetism with different ordering of magnetic moments, coexistence of superconductivity and magnetism, coexistence of superconducting and SDW/CDW phases etc. [2-7]. To increase the superconducting transition temperature $T_C$ in FeAs (FeSe) superconductors, doping of various atomic sublattices or application of external pressure are used [8-14]. There are phases with stoichiometric composition such as LiFeAs ($T_C$=18 K), LaFePO ($T_C$=6 K), FeSe ($T_C$=9 K), KFe$_2$As$_2$ ($T_C$=3.6 K) and others which exhibit superconducting properties without such manipulations.

Recently, among these compounds a family of new layered superconductors CaAFe$_4$As$_4$ (A=K, Rb, Cs) was prepared with the highest $T_C$ ~ 35 K for CaRbFe$_4$As$_4$ [15]. Further it was found by ARPES and DFT methods that for CaKFe$_4$As$_4$ a multiband character of superconductivity is realized [16]. This phase demonstrates two-gap s$_\pm$ superconductivity, which is common of high-temperature superconductivity in FeAs based systems [17]. In [18] an attempt was made to explain the emergence of superconductivity in CaKFe$_4$As$_4$ in the framework of the four bands Eliashberg theory.

Besides, a number of experimental thermodynamic and transport data such as anisotropic electrical resistivity, elastoresistivity, thermoelectric power, Hall effect, magnetization, and specific heat were obtained for CaKFe$_4$As$_4$ [19]. These data are close to those found for the related Ba$_{1-x}$K$_x$Fe$_2$As$_2$ system [20], which has a different crystal structure containing Ba and K sheets alternatively stacked with Fe$_2$As$_2$ layers, wherein the space group of the superconductor changes from P4/mmm to I4/mmm. In [21], the band structure and electronic properties of this new system with uniform distribution of alkali and alkali earth atoms in each layer were investigated for Ba$_{0.5}$K$_{0.5}$Fe$_2$As$_2$. It is established that Ba$_{1-x}$K$_x$Fe$_2$As$_2$ lies at the border of magnetic instability and pairing interactions in this «1144» phase might involve magnetic or orbital fluctuations.

Thus, the discovery of the new superconductors makes the investigation of their properties an interesting and actual problem from the theoretical point of view. For this purpose the band calculation methods based on the density functional theory (DFT) are often used. Today, as far as we know, no theoretical estimates of the properties of the new superconductors are available.

In this Communication we present the results of *ab initio* calculations of the properties of synthesized CaAFe$_4$As$_4$ (A=K, Rb, Cs) systems. We have obtained and analyzed the optimized atomic positions and different geometric parameters, band structures, Fermi surface topology, as well as total and partial densities of electronic states. Besides, the Bader's analysis of atomic charges and the charge density maps were used to discuss the inter-atomic bonding of these new superconductors.

**II. Model and computational aspects.** The synthesized new phases CaAFe$_4$As$_4$ (A=K, Rb, Cs) have [15] a tetragonal structure (space group No. 123 *P*4/*mmm*) and can be described as layered blocks [Fe$_4$As$_4$] alternatively stacked with two single sheets formed by Ca ions and ions of alkaline metals A, see Fig. 1. The atoms in this structure are arranged in the following positions: Ca in *1a* (0; 0; 0), K in *1d* (0.5; 0.5; 0.5), Fe in *4i* (0; 0.5; $z_{Fe}$); As in *2g* (0; 0; $z_{As1}$) and *2h* (0.5; 0.5; $z_{As2}$). Thus, the As atoms are located in two non-equivalent positions and are denoted as As$_1$ and As$_2$. The crystal structure of CaAFe$_4$As$_4$ is determined by the lattice parameters (*a* and *c*) and three internal parameters $z_{Fe}$, $z_{As1}$ and $z_{As2}$.

All the calculations were carried out by means of the full-potential linearized augmented plane wave method (FLAPW) implemented in the WIEN2k software package [22]. The generalized gradient approximation (GGA) to correct representation of exchange-correlation potential in the well-known PBE form [23] was used. The basis set inside each muffin tin (MT) sphere was split into core and valence subsystems. The core states were calculated within the spherical part of the potential and were assumed to have a spherically symmetric charge density in MT spheres. The valence part of the potential was expanded into spherical harmonics to $l = 4$. The valence wave functions inside the spheres were expanded into a row to $l = 12$. We used the following atomic radii: 2.20 a.u. for Ca and A, 2.00 a.u. for Fe and 1.85 a.u. for As atoms. The plane-wave expansion was taken to $R_{MT} \times K_{MAX}$ equal to 7, and the *k* sampling with 14×14×4 *k*-points in the Brillouin zone was used. The self-consistent calculations convergence conditions were the difference in the total energy of the crystal which did not exceed 0.1 mRy and the forces on atoms which did not exceed 1 mRy/a.u. as calculated at consecutive iterations. The total and partial densities of states (DOSs) were obtained by the modified tetrahedron method [24], and the features of the interatomic interactions picture were visualized by means of charge density maps. To describe the atomic charges, the Bader's analysis [25] was used additionally.

**III. Results and discussion.** Structural optimization of atomic positions was carried out. For all the calculations we used the experimental parameters of the unit cell: *a*=3.866 Å; *c*=12.817 Å for CaKFe$_4$As$_4$, *a*=3.876 Å; *c*=13.104 Å for CaRbFe$_4$As$_4$ and *a*=3.891 Å; *c*=13.414 Å for CaCsFe$_4$As$_4$. The obtained optimization results are presented in Table 1. It is seen that the values of parameters $z_{Fe}$, $z_{As1}$ and $z_{As2}$ for CaKFe$_4$As$_4$ agree well with the experimental values.

On the basis of these data we have calculated some geometric parameters of the investigated FeAs based layered superconductors. The obtained results are shown in Table II together with the experimental values of parameters *a* and $T_C$.

At first, we have analyzed the so called structure factor, i.e. a possible dependence of $T_C$ on the unit cell parameter *a*. According to this statement, $T_C$ grows with decreasing *a*. It is seen from Table II that this relation does not work because of the maximum $T_C$ for $CaRbFe_4As_4$. Thus, the enhancement of superconductivity in $CaAFe_4As_4$ phases cannot be explained by the influence of the structure factor.

Further note another parameter – atomic distances between Fe and As atoms, $d_{Fe-As}$. We see that all values $d_{Fe-As}$~2.30 Å, in satisfactory agreement with other related FeAs based superconductors. At the same time, the obtained values are slightly smaller than the experimental parameters of $d_{Fe-As}$~2.40 Å [15] for $CaRbFe_4As_4$.

Besides, some other empirical correlations between the crystallographic parameters and $T_C$ in high-temperature Fe pnictides are found [26-29], which allow us to estimate the characteristic parameters of $CaAFe_4As_4$ as superconductors with high values of $T_C$ and to compare them with the optimal values that are most favorable for superconductivity.

There is a correlation, according to which the highest $T_C$ is obtained when the bond angles As-Fe-As, $\alpha_{As-Fe-As}$, in $FeAs_4$ tetrahedrons reach the ideal value of 109.47°, which corresponds to ideal tetrahedron [26]. In this case we are dealing with two values of $\alpha_{As-Fe-As}$ due to the distortion of $FeAs_4$ tetrahedrons, see Table II. The angles are sufficiently close to this optimal parameter: for example, $\alpha_{As-Fe-As}$ = 112.1° and 107.6° for basic superconductor $CaKFe_4As_4$. This is consistent with the possibility of emergence of high $T_C$ superconductivity in these systems.

Another correlation of $T_C$ comes from the anion height $h_{Fe-As}$, which represents a height of As ions above the Fe atoms plane in FeAs blocks [27-29]. For FeAs based systems, this correlation has a maximum near $h_{Fe-As}$=1.38 Å, see [27]. An appropriate explanation is that the parameter $h_{Fe-As}$ may be a switch between high-$T_C$ nodeless and low-$T_C$ nodal pairings in FeAs based superconductors [29]. It should be mentioned that our parameters are $h_{Fe-As}$ ~1.27 Å and ~1.24 Å, i.e. these values are close to the above mentioned critical value and coincide for many other related systems.

Thus, the observed high-temperature superconductivity correlates well with such parameters as $\alpha_{As-Fe-As}$ and $h_{Fe-As}$, which is typical of systems based on FeAs blocks. At the same time, these experimental parameters for $CaRbFe_4As_4$ are much closer to the optimum in comparison with our calculated values, Table II.

The near-Fermi band structures and total and partial densities of states (DOSs) of FeAs based $CaAFe_4As_4$ superconductors are depicted in Figs. II, III. We have limited ourselves to this narrow energetic interval from -6 eV to 2 eV since it is this region that determines the electronic properties and interatomic interactions for the new investigated superconductors. The main band widths of $CaAFe_4As_4$ are given in Table III.

According to the obtained data, the energy spectrum of basic $CaKFe_4As_4$ includes the following characteristic bands: two lowest quasi-core bands separated by forbidden gaps and located about -14.4 eV and -11.4 eV below the Fermi level ($E_F$), which are due to the K 4*p* and As 4*p* states, and the broad valence band extending up to the Fermi level with a total width at ~ 5.6 eV, Fig. II. The latter is divided into two different subbands. The first subband is formed by partly hybridized Fe $3d$ and As $4p$ states (centered at -3 eV), which are responsible for the covalent Fe-As interactions. The second subband above -2 eV is due mainly to the Fe $3d$ states and adjoins directly $E_F$. These states form metallic-like Fe-Fe interactions in the Fe plane belonging to $Fe_4As_4$ layered blocks. In the process of transition from $CaKFe_4As_4$ to $CaCsFe_4As_4$ a slight narrowing of the total valence band is observed, Table III. The difference is in the number of lower bands and their width. While for $CaCsFe_4As_4$ there are two lower bands with similar composition but radically different band widths, for $CaRbFe_4As_4$ these bands merge into one band located at the top of the low part of the energy spectrum.

From Figs. II, III it is seen that the Fermi level falls on the region with a sufficiently high value of electronic states density causing a metallic type of conductivity for $CaAFe_4As_4$. The calculated total density of states at the Fermi level $N(E_F)$ is equal to 8.56-8.87 states/eV/form.unit or 2.14-2.22 states/eV/atom, Table IV. The maximum value of $N(E_F)$ is obtained for $CaKFe_4As_4$, and the minimum – for $CaRbFe_4As_4$; thus, this change does not coincide with $T_C$ variation, see above. Note that these values of $N(E_F)$ are comparable with $N(E_F)$=2.76 states/eV/atom for $Ba_{0.5}K_{0.5}Fe_2As_2$ [20] having a similar atomic composition and arbitrary distribution of Ba and K atoms in alternating layers, but a different space group I4/mmm. The estimated values of $N(E_F)$ allow us to calculate within the free electron model the low-temperature heat capacity coefficients $\gamma=(\pi^2/3)N(E_F)k^2_B$ and the Pauli paramagnetic susceptibility $\chi = \mu_B^2 N(E_F)$, the corresponding data are summarized in Table IV. Note that these values of $\gamma$ as calculated for Fe atom are smaller than for other layered FeAs based superconductors such as $Sr_2ScFeAsO_3$ ($\gamma$ = 8.85 mJ/K$^2$·mole [30]) or $Sr_3Sc_2Fe_2As_2O_5$ oxides ($\gamma$ = 6.77 mJ/K$^2$·mole [31]).

To establish the presence of magnetic properties in the investigated superconductors, we can use the obtained values of $N(E_F)$ according to the Stoner's criterion: $N(E_F)_s·I > 1$, where $N(E_F)_s$ is the total spin density at the Fermi level and $I$ is the exchange parameter. For $3d$ metals this parameter is ~0.7 [32]. Thus, it is seen that this rule is not observed because the multiplication is equal only to ~0.8, and this means that no significant magnetic moments should not be expected on Fe atoms in FeAs based $CaAFe_4As_4$ phases. Therefore our consideration of the properties of $CaAFe_4As_4$ only in non-magnetic case, as noted above, is reasonable.

The region near the Fermi level is very important because it participates in the creation of superconducting state. The partial contributions to $N(E_F)$ are given in Table IV. The main contribution to the electronic density of states around $E_F$ is made by the Fe $3d$ states that are the primary cause of emergence of metallic-like properties of $CaAFe_4As_4$ phases. The Fe $3d$ states give ~ 77 % of the overall contribution to $N(E_F)$. An appreciable contribution to $N(E_F)$ is also made by the As $4p$ and Fe $4s$, $4p$ states, Table IV. On the contrary, the contributions of K and Ca ions to the region near the Fermi level are very small, and therefore the conduction in this layered material must be highly anisotropic.

All the five Fe 3*d* orbitals contribute to the electronic density of states in the near-Fermi level region, see Fig. II and Table V. These orbitals give the following contribution to $N(E_F)$, in descending order: $3d_{x^2-y^2}$, $3d_{xz}$, $3d_{yz}$, $3d_{z^2}$, $3d_{xy}$. This sequence can be explained by the spatial arrangement of Fe *d* orbitals in accordance with the crystal structure of CaAFe$_4$As$_4$ superconductors. Thus, for example the $3d_{x^2-y^2}$ orbital having the maximum value of Fe 3*d* partial DOS is directed to the closest Fe atoms in the Fe plane forming Fe-Fe metallic interactions and determining the metallic-like properties of the material. At the same time, the $3d_{xy}$ orbital is involved in the Fe-Fe interactions to a less degree and shifts from the Fermi level, Fig. II. The remaining $3d_{xz}$, $3d_{yz}$ and $3d_{z^2}$ orbitals affecting the z axis also make a considerable contribution to $N(E_F)$, Table V. It is also clear from Table V that for As orbitals the $p_z$ states make a greater contribution than the $p_x$ and $p_y$ states.

These Fe 3*d* orbitals for CaAFe$_4$As$_4$ phases participate in the formation of the near Fermi bands and determine the topology and the main properties of the Fermi surfaces (FS), Fig. III. It is seen that the FS of these new FeAs based superconductors have a multisheet character and consist of 11 different electron or hole surfaces. The first 4 surfaces are close to the perfect cylinders of electronic type and are located at the edges of the Brillouin zone along the A-M direction. The next 6 surfaces located near the Γ point are also cylinders, but of hole type, and the external cylinder is slightly distorted and has a cup like form. The last surface is a small hole pocket at the center of the Birllouin zone near the Γ point. We can conclude that the basic feature of CaAFe$_4$As$_4$ FS is the predominance of cylindrical sheets, which is typical of other FeAs based layered superconductors, for example, LiFeAs [33], Sr$_2$ScFeAsO$_3$ [30] or Sr$_3$Sc$_2$Fe$_2$As$_2$O$_5$ [31]. Note that the calculated FS for CaKFe$_4$As$_4$ coincides well with that obtained by the LDA method [16].

As obtained from our FLAPW-GGA calculations, the interatomic interactions in the layered CaAFe$_4$As$_4$ compounds may be considered as a complex of anisotropic metallic, ionic and covalent bonding. We discuss the interactions in these new superconductors on the example of CaKFe$_4$As$_4$ using the charge density distribution of valence states in the (100) and (110) planes, see Fig. IV. In the Fe-As layers, first of all metallic Fe-Fe interactions are implemented between neighboring Fe atoms defining the properties of these superconductors. Besides, Fig. IV clearly shows the presence of covalent Fe-As interactions that are due to Fe 3*d* – As 4*p* hybridization, see the corresponding DOSs in Fig. II. Additionally, some Fe-As ionic interaction takes place because of different electronegativity of Fe and As atoms. Besides, a very small covalent bonding of As-As type also occurs in these Fe$_4$As$_4$ blocks that can be seen in Fig. IV. So, in the superconducting Fe$_4$As$_4$ layers there are all three basic types of interactions. Simultaneously, between the layers and Ca, A sheets mainly the ionic interaction dominates, and there must be a significant ion transport from alkali and alkali earth metals to As atoms belonging to $Fe_4As_4$ layers.

We have checked this assumption by estimating the atomic charges using the well-known Bader scheme [25]. The calculated values of $Q$ together with the charges of neutral atoms and the charges in case of a simple ionic model are presented in Table VI. It is seen that the results for Ca and alkali atom $A$ are almost the same as in the ionic model, thus confirming predominantly ionic bonds between Ca, A sheets and FeAs blocks. For Fe atoms the calculated atomic charges are about 0, which is typical of atoms with metallic interactions. This fact is an additional confirmation of the importance of metallic-like Fe-Fe interaction in the new synthesized superconductors. Finally, the charges of As atoms are negative, but are much greater than in case of a purely ionic model owing to covalent Fe-As interactions in the metallic blocks. In addition, it is worth noting that the difference in $Q$ for As atoms in non-equivalent positions 1 and 2 reaches ~0.2e and the relation $Q_{As1}<Q_{As2}$ works, Table VI. The charge transfer between Ca, A ions to FeAs blocks is calculated to be ~(2.1-2.2) e.

For very important metallic bonding estimation we calculated the so called parameter of metallicity, $f_m = 0.026N(E_F)/n_e$, where $n_e = N/V$; N is the total number of valence electrons, and $V$ is the cell volume [34]. The calculated values of $f_m$ are 0.163 for $CaKFe_4As_4$, 0.168 for $CaRbFe_4As_4$ and 0.171 for $CaCsFe_4As_4$, which is not very high. For example, the given values are smaller than those for other FeAs based superconducting oxide $Sr_4Sc_2Fe_2As_2O_6$, $f_m=0.36$ [35, 36].

**IV. Conclusions.** In summary, within the first-principles FLAPW-GGA calculations, we studied in detail the band structures, electronic properties, Fermi surface topology, and the interatomic bonding picture for the recently synthesized layered high-temperature superconductors $CaAFe_4As_4$ (A=K, Rb, Cs) with $T_C$ ~ 30 K. The superconductivity in these $CaAFe_4As_4$ compounds as in other related systems correlates well with such structure parameters as bond angles and anion height taking the values close to the optimum.

The total densities of states at the Fermi level are rather high for the investigated new superconductors, ~8.6-8.9 states/eV/form.unit, which is rather close to the related phase $Ba_{0.5}K_{0.5}Fe_2As_2$. It is found that the near-Fermi electronic bands, which are responsible for the formation of superconducting state, are due mainly to the Fe $3d$ states in FeAs layers, as in other FeAs superconductors. Moreover, among all Fe $3d$ states, the $d_{x2-y2}$, $d_{xz}$, and $d_{yz}$ orbitals give the maximum contribution to the near Fermi region and determine the Fermi surface features. The latter is of a multi-sheet type and combines mainly electronic and hole cylinder like surfaces at the edges and around the center of the Brillouin zone. This property is typical of FeAs based layered superconductors and is indicative again of possible outstanding superconducting properties of $CaAFe_4As_4$.

The inter-atomic bonding in $CaAFe_4As_4$ is highly anisotropic and includes a mixture of covalent, ionic, and metallic contributions inside $Fe_4As_4$ blocks, whereas the ionic bonding is realized between $Fe_4As_4$ blocks and atomic Ca and A sheets. By calculating the atomic charges $Q$ within the Bader scheme we confirmed the ionic interaction between these layers, whereas the values of $Q$ in $Fe_4As_4$ block are far from those for the purely ionic model. Thus, the atomic charges on Fe atoms are close to zero, which points to the presence of Fe-Fe metallic-like interactions between neighboring Fe atoms in $Fe_4As_4$ layers defining the main properties of the superconductors. We also estimated the charge transfer from Ca atom and alkali atom A to $Fe_4As_4$ blocks to be ~(2.1-2.2) e.

## V. Gratitude

The study was supported by the Russian Academy of Sciences (project No. AAAA-A16-116122810214-9).

## References

1. Y. Kamihara, T. Watanabe, M. Hirano, H. Hosono, J. Am. Chem. Soc. **130**, 3296 (2008).
2. A. L. Ivanovskii, Physics - Uspekhi **51**, 1229 (2008).
3. Z. A. Ren, Z. X. Zhao, Adv. Mater. **21**, 4584 (2009).
4. J. Paglione, R .L. Greene, Nature Phys. **6**, 645 (2010).
5. A.L. Ivanovskii, Russ. Chem. Rev. **79**, 1 (2010).
6. D. C. Johnson, Adv. Phys. **59**, 803 (2010).
7. D. Johrendt, J. Mater. Chem. **21**, 13726 (2011).
8. H. Okada, K. Igawa, H. Takahashi, Y. Kamihara, M. Hirano, H. Hosono, K. Matsubayashi, Y. Uwatoko, J. Phys. Chem. Soc. **77**, 113712 (2008).
9. P.M. Shirage, K. Miyazawa, H. Kito, H. Eisaki, A.Iyo, Phys. Rev. B **78**, 172503 (2008).
10. H. Takahashi, H. Okada, K. Igawa, K. Arii, Y. Kamihara, S. Matsuishi, M. Hirano, H. Hosono, K. Matsubayashi, Y. Uwatoko, J. Phys. Chem. Soc. **77**, 78 (2008).
11. K. Miyazawa, S. Ishida, K. Kihou, P.M. Shirage, M. Nakajima, C.H. Lee, H. Kito, Y. Tomioka, T. Ito, H. Eisaki, H. Yamashita, H. Mukuda, K. Tokiwa, S. Uchida, A. Iyo, Appl. Phys. Lett. **96,** 072514 (2010).
12. H. Takahashi, H. Soeda, M. Nukii, C. Kawashima, T. Nakanishi, S. Iimura, Y. Muraba, S**.** Matsuishi, H. Hosono, Sci. Reports **5**, 7829 (2015).
13. X.H. Chen, T. Wu, G. Wu, R.H. Liu, H. Chen, D.F. Fang, Nature **453**, 761 (2008).
14. G.F. Chen, Z. Li, G. Li, J. Zhou, D. Wu, J. Dong, W.Z. Hu, P. Zheng, Z.J. Chen, H.Q. Yuan, Phys, Rev. Lett. **101**, 057007 (2008).
15. A. Iyo, K. Kawashima, T. Kinjo, T. Nishio, S. Ishida, H. Fujihisa, Y. Gotoh, K. Kihu, H. Eisaki, Y, Yoshida, J. Amer. Chem. Soc. **138**, 3410 (2016).
16. D. Mou, T. Kong, W.R. Meier, F. Lochner, L.-L. Wang, Q. Lin, Y. Wu, S.L. Bud'ko, I. Eremin, D.D. Johnson, P.C. Canfield, A. Kaminski, Phys. Rev. Lett. **117**, 277001 (2016).
17. K. Cho, A. Fente, S. Teknowijoyo, M.A. Tanatar, K.R. Joshi, N.M. Nusran, T. Kong, W.R. Meier, U. Kalauarachchi, I. Guillamón, H. Suderow, S.L. Bud'ko, P.C. Canfield, R. Prozorov, Phys. Rev. B **95**, 100502(R) (2017).
18. G.A. Ummarino, Physica C **529**, 50 (2016).
19. W.R. Meier, T. Kong, U.S. Kaluarachchi, V. Taufour, N.H. Jo, G.Drachuck, A.E. Böhmer, S.M. Saunders, A. Sapkota, A. Kreyssig, M.A. Tanatar, R. Prozorov, A.I.

Goldman, F.F. Balakirev, A. Gurevich, S.L. Bud'ko, P.C. Canfield, Phys. Rev. B **94**, 064501 (2016).
20. M. Rotter, M. Tegel, D. Johrendt, Phys. Rev. Lett. **101**, 107006 (2008).
21. I.R. Shein, A.L. Ivanovskii, JETP Lett. **88**, 115 (2008).
22. P. Blaha, K, Schwarz, G.K.H. Madsen, D. Kvasnicka, J. Luitz, *WIEN2k, An Augmented Plane Wave Plus Local Orbitals Program for Calculating Crystal Properties*, Vienna University of Technology, Vienna, 2001.
23. J. P. Perdew, K. Burke, M. Ernzerhof, Phys. Rev. Lett. **77**, 3865 (1996).
24. P. E. Blöchl, O. Jepsen, O.K. Anderson, Phys. Rev. B **49**, 16223 (1994).
25. R. F. W. Bader, *Atoms in Molecules: A Quantum Theory, International Series of Monographs on Chemistry*, Clarendon Press, Oxford, 1990.
26. J.Zhao, Q.Huang, C. de la Cruz, S. Li, J.W. Lynn, Y. Chen, M.A. Green, G.F. Chen, G. Li, Z.C. Li, J.L. Luo, N.L. Wang, P. Dai, Nat. Mater. **7**, 953 (2008).
27. Y. Mizuguchi, Y. Hara, K. Deguchi, S. Tsuda, T. Yamaguchi, K. Takeda, H. Kotegawa, H. Tou, Y. Takano, Supercond. Sci. Technol. **23**, 054013 (2010).
28. E.Z. Kuchinskii, I.A. Nekrasov, M.V. Sadovskii, JETP Lett. **91**, 518 (2010).
29. K.Kuroki, H. Usui, S. Onari, R. Arita, H. Aoki, Phys. Rev. B **79**, 224511 (2009).
30. I. R. Shein, A. L. Ivanovskii, Phys. Rev. B **79**, 245115 (2009).
31. I. R. Shein, A. L. Ivanovskii, JETP Lett. **89**, 41 (2009).
32. T. Beuerle, K. Hummler, C. Elaasser, M. Faahnle, Phys. Rev. B **49**, 8802 (1994).
33. I.R. Shein, A.L. Ivanovskii, JETP Lett. **5**, 377 (2008).
34. Y. Li, Y. Gao, B. Xiao, T. Min, Z. Fan, S. Ma, L. Xu, J. Alloys Comp. **502**, 28 (2010).
35. I. R. Shein, A. L. Ivanovskii, JETP Lett. **89**, 41 (2009).
36. I. R. Shein, A. L. Ivanovskii, J. Supercond. Novel Maget. **22**, 613 (2009).

**TABLES**

Table I. The optimized atomic parameters ($z_{Fe}$, $z_{As1}$, $z_{As2}$) for new superconductors CaAFe$_4$As$_4$ (A=K, Rb, Cs) from FLAPW-GGA calculations.

| system | $z_{Fe}$ | $z_{As1}$ | $z_{As2}$ |
|---|---|---|---|
| CaKFe$_4$As$_4$ | 0.77787 | 0.32363 | 0.12471 |
| CaRbFe$_4$As$_4$ | 0.77803 (0.7754*) | 0.31893 (0.3336) | 0.12745 (0.1193) |
| CaCsFe$_4$As$_4$ | 0.77899 | 0.31637 | 0.12957 |

* experimental values for CaRbFe$_4$As$_4$ [15] are given in parenthesis.

Table II. The characteristic bond length ($d_{\text{Fe-As}}$, in Å), As anion height above the Fe atoms plane ($h_{\text{Fe-As}}$, in Å), bond angles ($\alpha_{\text{As-Fe-As}}$, in degrees) in comparison with the unit cell parameter ($a$, in Å) and superconducting transition temperature ($T_C$, in K) for CaAFe$_4$As$_4$ (A=K, Rb, Cs) superconductors.

| system | $d_{\text{Fe-As}}$ | $h_{\text{Fe-As}}$ | $\alpha_{\text{As-Fe-As}}$ | $a$ | $T_C$ |
|---|---|---|---|---|---|
| CaKFe$_4$As$_4$ | 2.3300; 2.3012 | 1.30106; 1.24850 | 112.12; 107.64 | 3.866* [15] | 33.1* [15] |

| | | | | | | |
|---|---|---|---|---|---|---|
| CaRbFe$_4$As$_4$ | 2.3172; 2.2999 (2.408; 2.380 [15]) | 1.27056; 1.23859 (1.429; 1.381) | 113.50; 107.76 (109.1; 107.2) | 3.876 [15] | 35.0 [15] | |
| CaCsFe$_4$As$_4$ | 2.3284; 2.2999 | 1.27916; 1.22657 | 113.35; 107.04 | 3.891 [15] | 31.6 [15] | |

\* experimental values [15] are given in parentheses.

Table III. The main band widths for CaAFe$_4$As$_4$ (A=K, Rb, Cs) superconductors as obtained from FLAPW-GGA calculations.

| system/type of band | A p band | Band gap 1 | A p + As 4p band | Band gap 2 | The valence band (up to $E_F$) |
|---|---|---|---|---|---|
| CaKFe$_4$As$_4$ | 0.52 | 1.54 | 2.41 | 4.68 | 5.55 |
| CaRbFe$_4$As$_4$ | - | - | 2.38 | 4.82 | 5.48 |
| CaCsFe$_4$As$_4$ | 2.26 | 0.41 | 1.43 | 3.06 | 5.37 |

Table IV. The total and partial densities of states at the Fermi level ($N_{tot}(E_F)$, $N^l(E_F)$, in states/eV/form.unit), Sommerfeldt coefficients ($\gamma$, in mJ/(K$^2$·mole) and molar Pauli paramagnetic susceptibility ($\chi$, in $10^{-4}$ emu/mole) for CaAFe$_4$As$_4$ (A=K, Rb, Cs) superconductors.

| system | $N(E_F)$ | $N^{Fe\ 3d}(E_F)$ | $N^{As\ 4p}(E_F)$ | $N^{Fe\ 4s+4p}(E_F)$ | $N^{Ca\ s,p\ +\ K\ s,p}(E_F)$ | $\gamma$ | $\chi$ |
|---|---|---|---|---|---|---|---|
| CaKFe$_4$As$_4$ | 8.870 | 6.903 | 0.095 (1) 0.098 (2) 0.193 | 0.096 | 0.010 | 21.01 | 2.72 |
| CaRbFe$_4$As$_4$ | 8.556 | 6.591 | 0.091 (1) 0.098 (2) | 0.095 | 0.012 | 20.26 | 2.63 |

| | | | 0.189 | | | | |
|---|---|---|---|---|---|---|---|
| CaCsFe$_4$As$_4$ | 8.731 | 6.688 | 0.093 (1)  0.096 (2)  0.189 | 0.097 | 0.016 | 20.68 | 2.68 |

Table V. The contribution of Fe d and As p orbitals to the electronic density of states at the Fermi level ($N^l$ ($E_F$), in states/eV/form.unit) for CaAFe$_4$As$_4$ (A=K, Rb, Cs).

| system | $N^{Fe\ dz2}$ ($E_F$) | $N^{Fe\ dx2-y2}$ ($E_F$) | $N^{Fe\ dxy}$ ($E_F$) | $N^{Fe\ dxz}$ ($E_F$) | $N^{Fe\ dyz}$ ($E_F$) | $N^{As\ pz}$ ($E_F$) | $N^{As\ px+py}$ ($E_F$) |
|---|---|---|---|---|---|---|---|
| CaKFe$_4$As$_4$ | 0.914 | 2.310 | 0.380 | 1.966 | 1.333 | 0.061 (1)  0.078 (2)  0.139 | 0.034 (1)  0.020 (2)  0.054 |
| CaRbFe$_4$As$_4$ | 0.956 | 2.176 | 0.387 | 1.937 | 1.135 | 0.060 (1)  0.079 (2)  0.139 | 0.031 (1)  0.019 (2)  0.050 |
| CaCsFe$_4$As$_4$ | 0.995 | 2.202 | 0.412 | 1.827 | 1.252 | 0.063 (1)  0.078 (2)  0.141 | 0.030 (1)  0.018 (2)  0.048 |

Table VI. The atomic charges calculated in the Bader model ($Q_B$, in e) for CaAFe$_4$As$_4$ (A=K, Rb, Cs) in comparison with neutral atoms (Z, in e), their difference ($\Delta Q = Z - Q_B$) and atomic charges in the purely ionic model ($Q_i$, in e).

| system | atom | Q | Z | ΔQ | $Q_i$ |
|---|---|---|---|---|---|
| $CaKFe_4As_4$ | Ca | 18.607 | 20 | +1.393 | +2 |
| | K | 18.216 | 19 | +0.784 | +1 |
| | Fe | 25.847 | 26 | +0.153 | +1.75 |
| | $As_1$ | 33.596 | 33 | -0.596 | -3 |
| | $As_2$ | 33.786 | 33 | -0.786 | -3 |
| $CaRbFe_4As_4$ | Ca | 18.594 | 20 | +1.406 | +2 |
| | Rb | 36.248 | 37 | +0.752 | +1 |
| | Fe | 25.839 | 26 | +0.161 | +1.75 |
| | $As_1$ | 33.582 | 33 | -0.582 | -3 |
| | $As_2$ | 33.807 | 33 | -0.807 | -3 |
| $CaCsFe_4As_4$ | Ca | 18.582 | 20 | +1.418 | +2 |
| | Cs | 54.298 | 55 | +0.702 | +1 |
| | Fe | 25.834 | 26 | +0.166 | +1.75 |
| | $As_1$ | 33.557 | 33 | -0.557 | -3 |
| | $As_2$ | 33.822 | 33 | -0.822 | -3 |

**FIGURES**

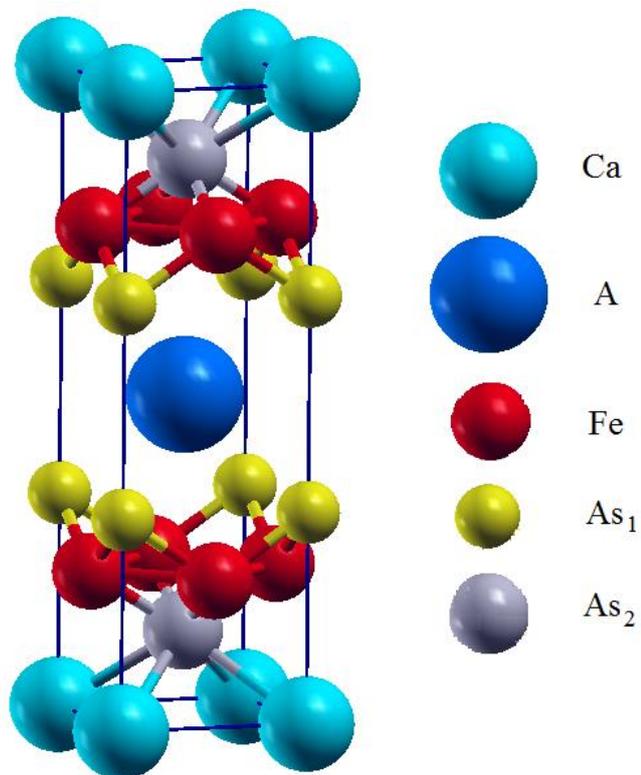

Fig. I. The crystal structure of the new high-temperature superconductors CaAFe$_4$As$_4$ (A=K, Rb, Cs) used in the FLAPW-GGA calculations.

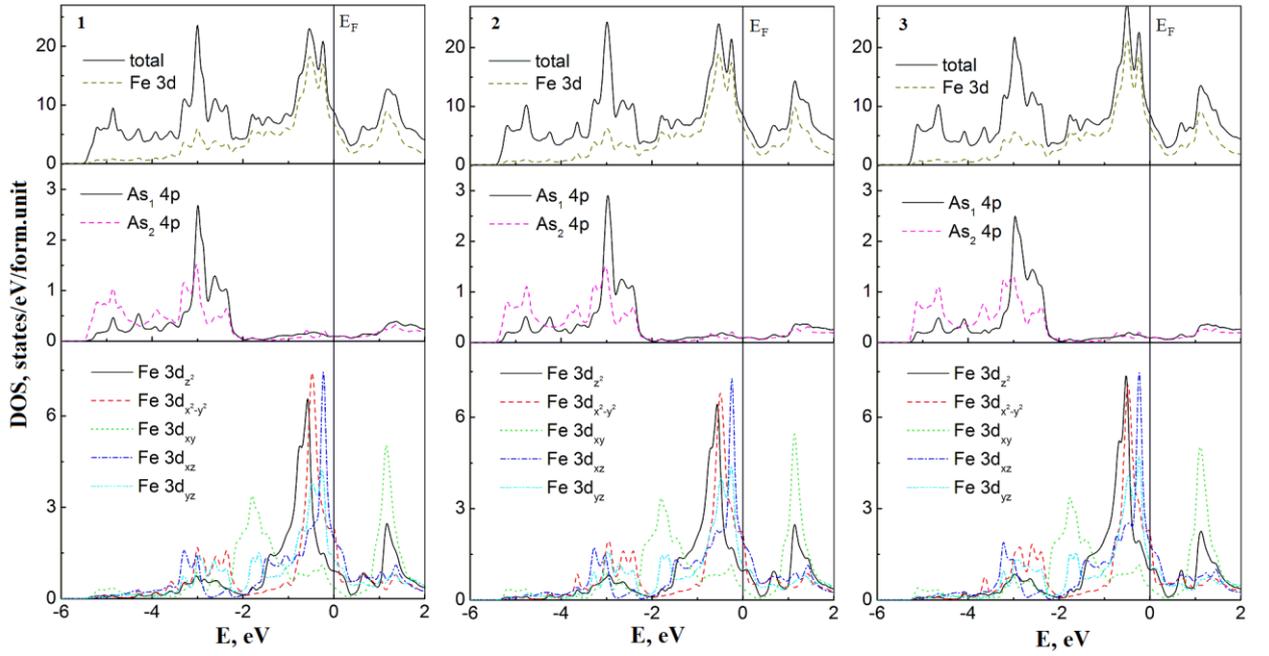

Fig. II. Total (above) and partial densities of electronic states for FeAs based superconductors $CaKFe_4As_4$ (1), $CaRbFe_4As_4$ (2) and $CaCsFe_4As_4$ (3).

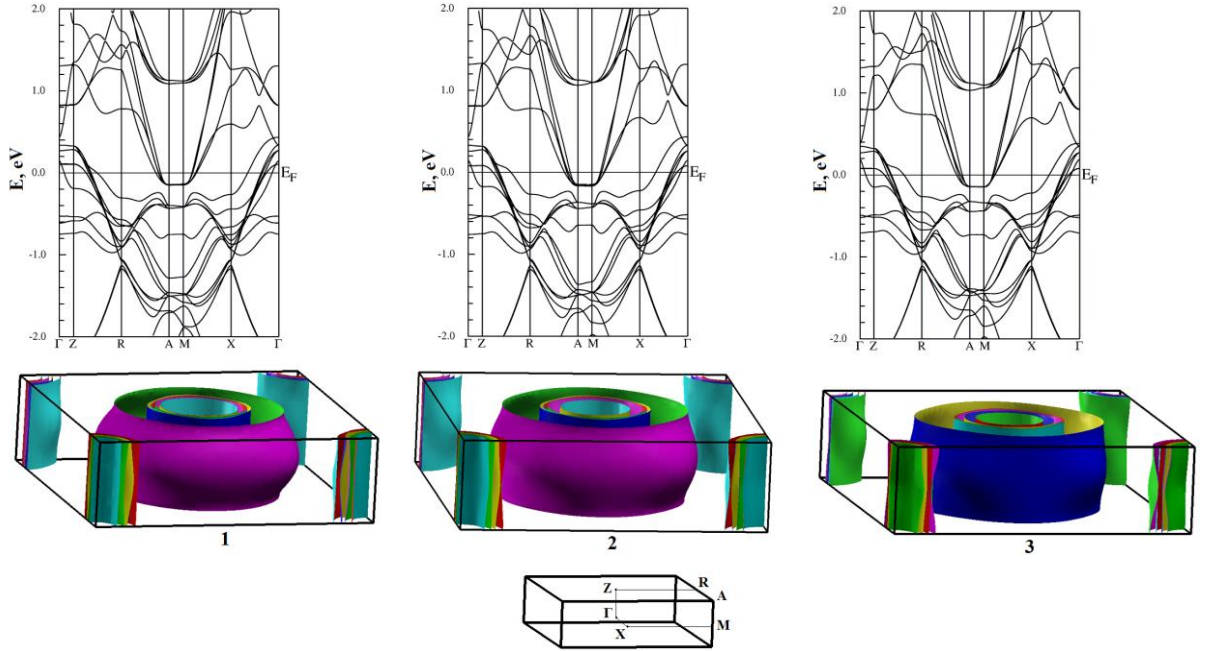

Fig. III. *Top* - the band structures in the near Fermi region for $CaKFe_4As_4$, $CaRbFe_4As_4$ and $CaCsFe_4As_4$; *bottom* - the corresponding Fermi surfaces for the new superconductors $CaKFe_4As_4$ (1), $CaRbFe_4As_4$ (2) and $CaCsFe_4As_4$ (3). The corresponding Brillouin zone is presented.

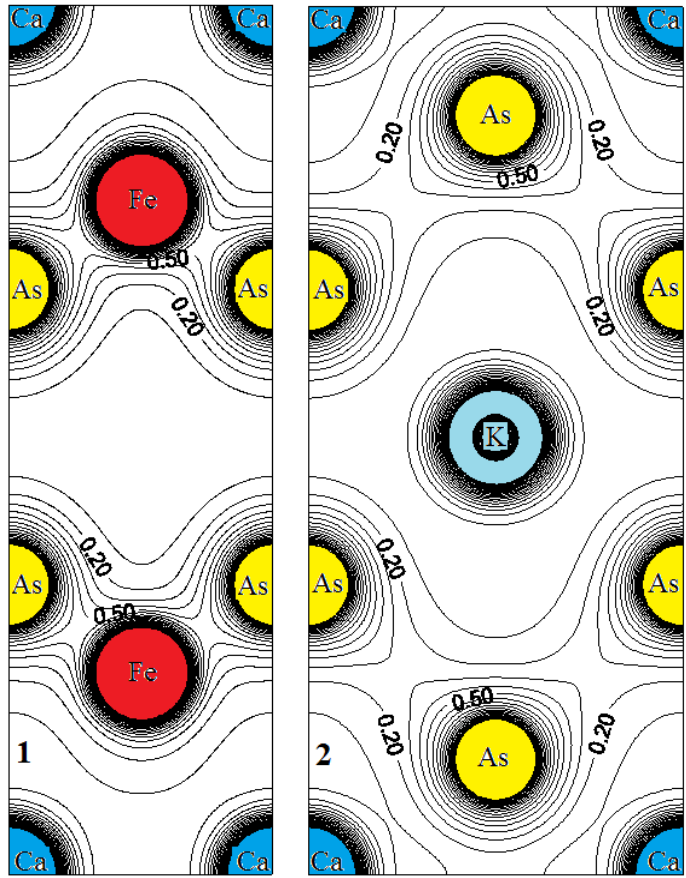

Fig. IV. The charge density distribution of the valence states in (100) and (110) planes for the basic superconductor $CaKFe_4As_4$ (1, 2). The interval between the isoelectronic contours is 0.1 e/Å$^3$.